\documentclass[conference]{IEEEtran}
\IEEEoverridecommandlockouts
\usepackage{cite}
\usepackage{amsmath,amssymb,amsfonts}
\usepackage{graphicx}
\usepackage{textcomp}
\usepackage{comment}
\usepackage{booktabs}
\usepackage{caption}
\usepackage{balance}

\usepackage{xcolor}
\usepackage{algorithm}
\usepackage{threeparttable}
 \usepackage{array} 
 
\usepackage{algpseudocode}

\def\BibTeX{{\rm B\kern-.05em{\sc i\kern-.025em b}\kern-.08em
    T\kern-.1667em\lower.7ex\hbox{E}\kern-.125emX}}
\begin{document}

\title{DaggerFFT: A Distributed FFT Framework Using Task Scheduling in Julia}
%DaggerFFT: A Task Scheduling Approach to Distributed FFTs in Julia for Scalable CPU/GPU Execution

\author{
\IEEEauthorblockN{
Sana Taghipour Anvari\IEEEauthorrefmark{1},
Julian Samaroo\IEEEauthorrefmark{2},
Matin Raayai Ardakani\IEEEauthorrefmark{1},
David Kaeli\IEEEauthorrefmark{1}
}
\IEEEauthorblockA{
\IEEEauthorrefmark{1}Northeastern University \quad
\IEEEauthorrefmark{2}Massachusetts Institute of Technology\\
\{taghipouranvari.s, raayaiardakani.m, d.kaeli\}@northeastern.edu \quad
jsamaroo@mit.edu
}
}

\maketitle
\begin{abstract}
The Fast Fourier Transform (FFT) is a fundamental numerical technique with widespread application in a range of scientific problems. As scientific simulations attempt to exploit exascale systems, there has been a growing demand for distributed FFT algorithms that can effectively utilize modern heterogeneous high-performance computing (HPC) systems. Conventional FFT algorithms commonly encounter performance bottlenecks, especially when run on heterogeneous platforms. Most distributed FFT approaches rely on static task distribution and require synchronization barriers, limiting scalability and impacting overall resource utilization.

%We present DaggerFFT a distributed FFT framework written entirely in Julia, that redefines highly parallel FFT computations as a dynamically scheduled task graph. Each FFT stage operates on a separately defined distributed array. FFT operations are expressed as DTasks operating on pencil or slab partitioned DArrays. Each FFT stage owns its own DArray, and the runtime schedules ready DTasks across devices using Dagger’s dynamic scheduler with work stealing. This design enables fine-grained task scheduling, backend-specific task placement, and improved overlap between communication and computation, while retaining flexibility for CPU and GPU execution. To address the data transfer bottleneck in distributed FFTs, DaggerFFT employs an asynchronous exchange with full overlap across five stages of each redistribution: posting receives, packing, nonblocking sends, local GPU-GPU copies, and progressive unpacking. This overlap reduces global synchronizations and ensures that communication proceeds in a pipelined order rather than as a sequence of blocking steps.

%Our results demonstrate that high-level languages, such as Julia, can achieve performance close to, and in many cases better than, traditional low-level HPC libraries by enabling flexible and modular algorithm design. 
In this paper we present DaggerFFT, a distributed FFT framework, developed in Julia, that treats highly parallel FFT computations as a dynamically scheduled task graph. Each FFT stage operates on a separately defined distributed array. FFT operations are expressed as DTasks operating on pencil or slab partitioned DArrays. Each FFT stage owns its own DArray, and the runtime assigns DTasks across devices using Dagger’s dynamic scheduler that uses work stealing.
We demonstrate how DaggerFFT's dynamic scheduler can outperform state-of-the-art distributed FFT libraries on both CPU and GPU backends, achieving up to a 2.6× speedup on CPU clusters and up to a 1.35× speedup on GPU clusters. We have integrated DaggerFFT into Oceananigans.jl, a geophysical fluid dynamics framework, demonstrating that high-level, task-based runtimes can deliver both superior performance and modularity in large-scale, real-world simulations.
\end{abstract}


\section{Introduction}
FFTs are fundamental numerical tools with applications across diverse fields, including engineering, physics, signal processing, and applied mathematics\cite{b9}.  
Despite decades of research and the availability of high-performance libraries, distributed FFTs remain difficult to scale efficiently. The challenges stem not only from the high communication costs of multi-dimensional transforms, but also from the rigidity of existing implementations. Most distributed FFT libraries rely on slab (1D) or pencil (2D) decompositions that assign fixed portions of the domain to each process. While effective under ideal conditions, these static decompositions lack flexibility; every stage of the transform applies the same partitioning, task placement is predetermined, and execution proceeds in tightly synchronized steps. This rigidity leads to inefficiencies such as unnecessary synchronization and suboptimal utilization of available hardware resources.

To address these limitations, we reformulate FFT execution around runtime scheduling rather than static decomposition. Each FFT stage is expressed as a collection of independent one-dimensional (or two-dimensional) transforms that are dynamically scheduled at runtime based on data dependencies and processor availability. Communication between stages is performed by a separate MPI-based pipeline with asynchronous, backend-aware transfers. This separation enables dynamic scheduling to improve utilization, reduce synchronization costs, and adapt naturally to heterogeneous or imbalanced systems, while communication remains efficient and predictable.

This paper introduces \textit{DaggerFFT}, a distributed 3D FFT framework that implements this design. Each transform stage operates on an independently declared distributed array (\texttt{DArray}), where the local FFT computations are expressed as distributed tasks (\texttt{DTasks}) scheduled dynamically by runtime. This task-based formulation enables stage-specific decompositions and fine-grained control over data locality. The result is a modular FFT framework that reduces synchronization, exploits locality, and achieves high utilization across diverse hardware configurations. Our framework is implemented in Julia programming language~\cite{b15}, which combines high-level expressiveness with performance through just-in-time compilation. Julia is increasingly being adopted in high-performance simulation codes\cite{b17} for demanding applications such as \textit{Oceananigans.jl}~\cite{b27} for ocean modeling and \textit{GeophysicalFlows.jl}~\cite{b26} for geophysical fluid dynamics simulations, both of which rely heavily on FFT-based solvers.

This work makes the following contributions:
\begin{itemize}
    \item We implement a distributed FFT framework that integrates a task-based scheduling runtime for transformations with an independent MPI communication layer, demonstrating how high-level task frameworks can accelerate computation-intensive HPC kernels.
    
    \item We express Fourier transforms using distributed tasks (\texttt{DTasks}) operating on distributed arrays (\texttt{DArray}), enabling fully asynchronous execution of local transforms. Task dependencies are managed automatically by the runtime, which allows independent FFT stages to proceed without global synchronization while still supporting optional work stealing for improved utilization.
    
    \item We design a communication layer that performs data redistribution between \texttt{DArrays} using non-blocking collectives and efficient buffer management. This pipeline overlaps packing, transfer, and unpacking operations to reduce synchronization overhead.
    
    \item We present a performance evaluation demonstrating that \textit{DaggerFFT} achieves competitive or superior performance compared to state-of-the-art distributed FFT libraries implemented in low-level languages, while offering greater flexibility and a user-friendly interface for HPC applications.
    
    \item We rewrite the Distributed Poisson solver in \textit{Oceananigans.jl}~\cite{b27} using \textit{DaggerFFT}, demonstrating that our framework can be integrated with minimal modifications into production-scale geophysical fluid dynamics models and deliver tangible performance improvements.
\end{itemize}

\begin{comment}
Our implementation supports slab and pencil decompositions, 1D/2D/3D FFTs, different types of forward and backward transforms, and runtime backend selection across CPUs and GPUs. Performance evaluations demonstrate that our framework can achieve comparable or superior runtimes to state-of-the-art MPI-based distributed FFT libraries, with particularly strong performance on GPUs. At the same time, our library offers significantly greater modularity and adaptability. As a result, this work demonstrates that a high-level dynamic task scheduling approach can effectively support distributed FFT computations, delivering strong performance without sacrificing flexibility or portability.
\end{comment}

\section{Related Work and Distributed FFT Fundamentals}
Fast Fourier Transform (FFT) libraries have undergone decades of optimization across both single-node and distributed environments. On single-node systems, libraries such as FFTW~\cite{b1}, Intel~MKL~\cite{b10}, cuFFT~\cite{b11}, and rocFFT~\cite{b12} remain the established standards, providing highly tuned implementations for CPUs and GPUs through platform-specific vectorization, memory planning, and autotuning. These packages serve as the computational backbone for many scientific applications, though they are constrained by the memory capacity of a single node.

\subsection{Distributed FFT Libraries on CPUs}

To overcome single-node limitations, distributed FFT libraries extend these kernels to multi-node environments using domain decomposition. Early efforts such as FFTW-MPI~\cite{b1} and FFTMPI~\cite{b3} use slab decomposition, partitioning data along one dimension so each process holds a full plane of the remaining dimensions. Slab-based approaches minimize communication during the first transform stage and achieve reliable performance on moderate process counts, but scalability is fundamentally limited because the number of slabs cannot exceed the size of one dimension~\cite{b25}.  

P3DFFT~\cite{b2} and 2DECOMP\&FFT~\cite{b25} employed pencil decomposition, which divides data along two dimensions to increase the number of partitions and improve load balancing. Pencil decomposition enables strong and weak scaling to thousands of ranks and is widely adopted in computational fluid dynamics and turbulence simulations~\cite{b25}. OpenFFT~\cite{b14} extends these ideas through automatic decomposition tuning and locality-aware communication strategies that adapt to different MPI topologies. Despite these advances, weak scaling in such libraries degrades at extreme process counts due to increased transpose traffic and global synchronization overheads. Other domain-specific frameworks, such as SWFFT~\cite{b4}, have focused on modularity and scalability for cosmological applications, while still relying on traditional slab or pencil decompositions.

\subsection{GPU-Accelerated and Hybrid Frameworks}

The rise of GPU-accelerated systems motivated distributed FFT libraries that exploit high bandwidth and massive parallelism. AccFFT~\cite{b6} generalizes pencil decomposition to hybrid MPI–CUDA environments, overlapping communication and computation to hide PCIe latency. heFFTe~\cite{b5} provides a portable, exascale-oriented framework supporting multiple backends, it employs MPI with OpenMP threading on CPUs and CUDA-aware MPI for GPU execution. Earlier GPU-focused libraries such as FFTE~\cite{b7} achieved high throughput on vendor-specific interconnects, but lacked portability. In comparison, heFFTe emphasizes open standards and asynchronous data packing to hide communication cost.  Across these efforts, the community has shifted from slab to pencil decompositions and adopted hybrid MPI/OpenMP execution~\cite{b13}, as well as non-blocking MPI collectives~\cite{b16}, to overlap computation and communication.  Nevertheless, most of these frameworks continue to use static decompositions and bulk-synchronous execution, limiting adaptability on heterogeneous clusters.

\subsection{Runtime and Architecture Aware FFT Studies}

Beyond decomposition-driven designs, several studies explore run-time level or architecture-specific formulations. GFFT~\cite{b28} expresses the FFT as a fine-grained dependency graph to enable flexible scheduling and fusion of operations, demonstrating improved single node concurrency. OpenFFT-SME~\cite{b29} leverages ARM’s Scalable Matrix Extension (SME) to accelerate butterfly operations. These projects illustrate different strategies for improving performance through dynamic scheduling and hardware-aware optimization, but typically target single node or hardware-specific environments.

\subsection{Positioning of This Work}

We reformulate the distributed FFT pipeline as a sequence of small, independent FFT tasks that are dynamically scheduled and executed asynchronously by a task-based runtime. Each transform stage operates on independently defined distributed arrays and allows stage-specific decompositions compatible with slab and pencil layouts. Inter-stage communication employs asynchronous, non-blocking MPI transfers with coalesced data packing, enabling bulk data exchanges that can be overlapped with computation and minimize synchronization. This design bridges traditional domain decomposition methods with modern task-based runtimes and enables a new class of distributed FFT implementations that can be efficiently implemented on heterogeneous systems.

\section{Problem Formulation and Runtime Model}
\subsection{Mathematical Structure of the 3D FFT}

A three-dimensional Fast Fourier Transform (3D FFT)\cite{b8} on an array $A \in \mathbb{C}^{N_x \times N_y \times N_z}$ is defined as the tensor product of three one-dimensional FFTs:
\[
\hat{A}(k_x, k_y, k_z)
= \sum_{i=0}^{N_x-1}\sum_{j=0}^{N_y-1}\sum_{l=0}^{N_z-1}
A(i,j,l)\,
e^{-2\pi i \left(\frac{k_x i}{N_x} + \frac{k_y j}{N_y} + \frac{k_z l}{N_z}\right)}.
\]
Operationally, the 3D FFT is evaluated as a sequence of three independent 1D transforms applied along the $x$, $y$, and $z$ dimensions. Between stages, data are redistributed to align the decomposition with the dimension being transformed.

The sequence for a two-dimensional (pencil) decomposition:
\begin{enumerate}
  \item Perform $N_y\times N_z$ independent 1D FFTs of length $N_x$;
  \item Redistribute data for the $y$-direction;
  \item Perform $N_x\times N_z$ independent 1D FFTs of length $N_y$;
  \item Redistribute data for the $z$-direction;
  \item Perform $N_x\times N_y$ independent 1D FFTs of length $N_z$.
\end{enumerate}

For a one-dimensional (slab) decomposition, the first two transforms are performed locally on each slab before a single global transpose redistributes data for the final transform along the remaining dimension.

Each redistribution corresponds to a global data transpose, whose cost is determined by the decomposition strategy and communication pattern. Although the algorithmic complexity of the FFT is $O(N\log N)$\cite{b1}, scalability in practice is dominated by data movement and synchronization across these stages.~\cite{b16}

\subsection{Distributed Execution Model}

\begin{figure*}[t]
    \vspace{-0.3cm}
    \centering
    \includegraphics[width=0.95\linewidth]{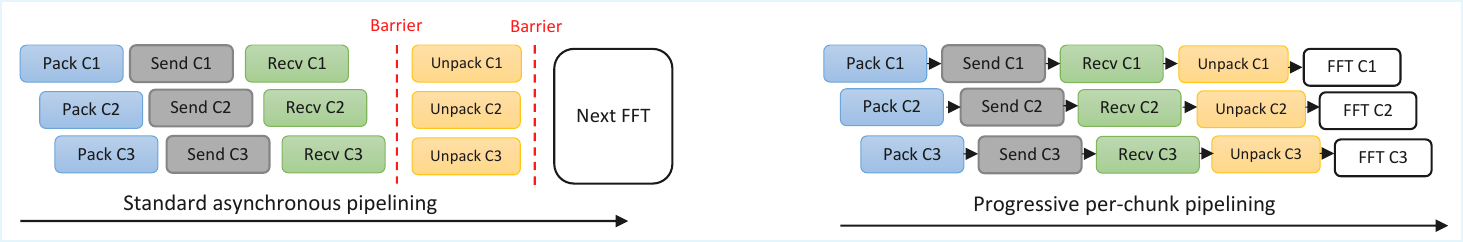}
    \vspace{-0.2cm}
    \caption{Comparison between standard asynchronous redistribution and the proposed progressive per-chunk pipelining. 
    In prior approaches (left), unpacking and subsequent FFTs begin only after all sends are complete, requiring implicit barriers. In our design (right), receives and unpacks occur progressively as messages arrive, enabling continuous overlap.}
    \label{fig:pipeline_overlap}
    \vspace{-0.5cm}
\end{figure*}

The redistribution phase in a distributed FFT aligns data partitions with the dimension being transformed in the next stage. This step dominates communication cost and involves three interdependent operations: i) packing local subdomains, ii) exchanging data among MPI ranks, and iii) unpacking received buffers. 
Existing distributed FFT libraries~\cite{b5} adopt asynchronous communication to overlap packing and data transfer, but still enforce explicit synchronization barriers, typically waiting for all sends to complete before beginning the unpack operation or the next computation phase.  As a result, there is only a limited amount of overlap between communication and local computation  (Figure~\ref{fig:pipeline_overlap}). The details of the algorithm are explained in Algorithm~\ref{alg:flat_redistribute}.

The total communication time per redistribution phase can be modeled using the following latency–bandwidth formulation:
\begin{equation}
\vspace{0.2cm}
T_{\text{comm}} \approx \alpha |S| + \beta m,
\vspace{-0.2cm}
\end{equation}

where $\alpha$ denotes message startup latency, $\beta$ the inverse bandwidth, $|S|$ the number of peers, and $m$ the total transferred data volume. 

In the ideal case of perfect overlap among packing, communication, and unpacking, 
the effective redistribution time approaches the lower bound
\begin{equation}
\vspace{0.2cm}
T_{\text{eff}} \gtrsim \max(T_{\text{pack}},\, T_{\text{mpi}},\, T_{\text{unpack}}),
\vspace{-0.2cm}
\end{equation}

representing the best achievable scenario, where all stages progress concurrently. In practice, dependencies, limited buffer space, and network contention prevent full concurrency, leading to partially overlapped execution. Our pipelined design minimizes these stalls by removing global synchronization, allowing continuous processing of packing, data transfers, and unpacking within the same phase.

\subsection{Runtime Scheduling Model}
\label{sec:runtime_model}
Each FFT stage generates a large number of fine-grained computation tasks, 
$t_i \in T$, each operating on a local data chunk $c_i$ distributed across processors 
$p_j \in P$. Task execution time depends on both the computation cost and the cost 
of accessing non-local data.  Following the LogP model of parallel computation 
introduced by Culler~\textit{et al.}~\cite{b31}, we express the placement cost of task $t_i$ on processor $p_j$ as
\begin{equation}
\vspace{0.2cm}
w_{i,j} = C_{\text{comp}}(t_i, p_j) + C_{\text{comm}}(t_i, p_j),
\vspace{-0.2cm}
\label{eq:scheduling_cost}
\end{equation}

where $C_{\text{comp}}$ is the estimated compute time on $p_j$ and
\begin{equation}
\vspace{0.2cm}
C_{\text{comm}}(t_i, p_j) = L + \frac{V}{B},
\vspace{-0.2cm}
\end{equation}
models the latency–bandwidth cost for transferring data volume $V$ across effective bandwidth $B$ with one-way latency $L$. Tasks are first scheduled on the processor that already holds their input chunk, ensuring data-local execution. When a processor’s local queue becomes empty, the runtime performs work stealing so that idle threads or ranks fetch remote data chunks only if their predicted idle time exceeds the communication cost.

The additional latency incurred by stealing a task $t_i$ from a remote queue can be modeled as
\begin{equation}
\vspace{0.2cm}
\tau_s(t_i) = L + \frac{V}{B} + \sigma,
\vspace{-0.2cm}
\end{equation}
where $\sigma$ represents runtime overhead due to queue management and task serialization. Workload stealing is only performed if the predicted idle time $I_q$ of the thief is longer than the steal cost, i.e.,
\begin{equation}
\vspace{0.2cm}
I_q > \tau_s(t_i).
\vspace{-0.3cm}
\end{equation}

This formulation ensures that work stealing improves utilization only when the communication latency can be amortized by the available idle time.

The per-phase runtime can thus be modeled as
\begin{equation}
\vspace{0.3cm}
T_{\text{phase}} \approx
\max(T_{\text{comp}},\, T_{\text{comm}}) + (1-\rho)\,k\tau_s,
\vspace{-0.3cm}
\end{equation}
where $T_{\text{comp}}$ represents the time for workers executing locally 
placed FFT tasks, $T_{\text{comm}}$ represents the cumulative time for 
workers fetching and executing stolen tasks, $k$ is the number of tasks 
per worker, $\tau_s$ is the per-task scheduling cost, and $\rho$ is the 
fraction of overhead hidden by parallelism. The $\max$ operator reflects 
that local and stolen work proceed in parallel across workers. As task 
granularity decreases (larger $k$), exposed scheduling overhead 
$(1-\rho)k\tau_s$ grows while per-task costs decrease.

In practice, most of the observed performance gains in our system arise not from aggressive stealing,
but from three complementary runtime properties:
\begin{itemize}
  \item \textbf{Asynchronous task spawning:} FFT tasks are created and scheduled without global barriers, reducing idle time and improving concurrency.
  \item \textbf{Memory reuse:} Persistent workspaces and buffer reuse reduce allocation overhead and improve cache and GPU-memory locality.
    \item \textbf{Locality-driven placement:} Tasks preferentially execute where their input data reside, minimizing data movement.
\end{itemize}

This model reflects the core trade-offs observed in our implementation. Local placement minimizes communication overhead, but may introduce imbalance, while dynamic work stealing helps restore balance when workloads or resources are uneven. By combining a representation of communication with adaptive task scheduling, the runtime sustains high utilization across heterogeneous systems. The benefits of dynamic scheduling are most evident in real world applications, especially when working with sparse or unevenly distributed data such as spectral solvers~\cite{b2}, particle simulations~\cite{b30}, or fluid models~\cite{b27}, where computational load varies across space and time. 

\section{Algorithm Design}
\subsection{Distributed 3D FFT Pipeline Integration}

\begin{figure*}[h]
\vspace{-0.5cm}
    \centering
    \includegraphics[width=\linewidth]{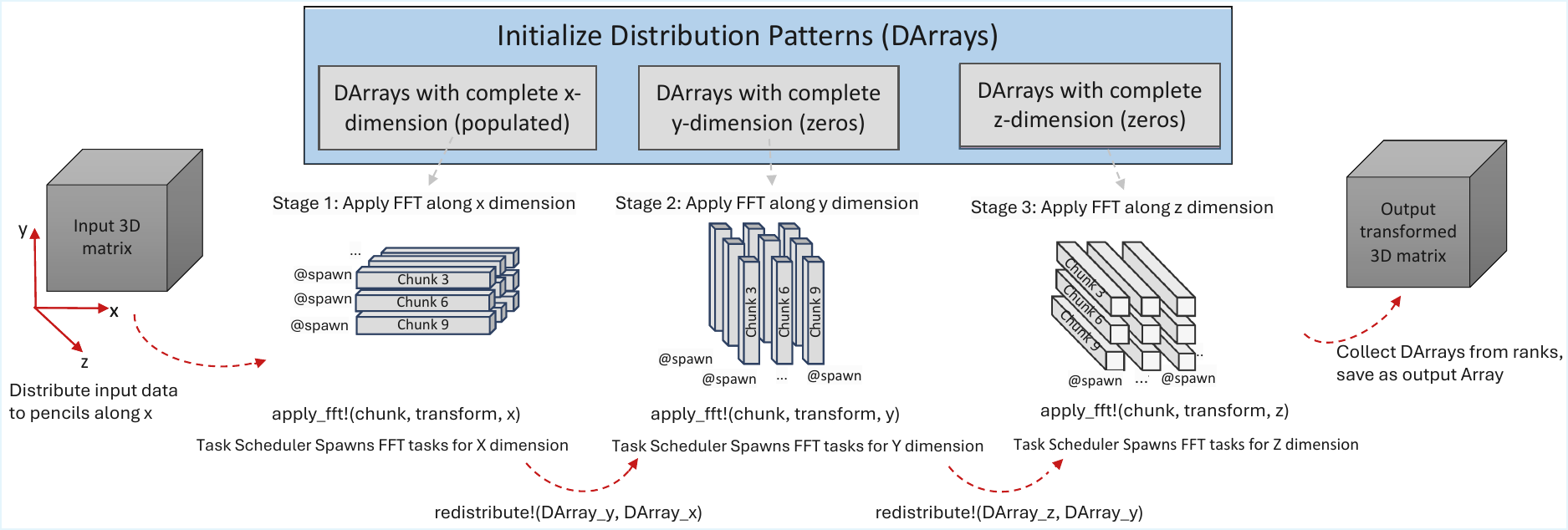}
    \vspace{-0.5cm}
    \caption{Task-scheduled 3D FFT implementation using pencil decomposition, asynchronous transforms and data movement.}
    \vspace{-0.5cm}
    \label{fig:decomposition}
    \vspace{-0.4cm}
\end{figure*}

A distributed 3D FFT consists of three FFT stages (x-, y-, and z-directions), separated by two data redistribution phases, as illustrated in 
Figure~\ref{fig:decomposition}. Each stage applies local one-dimensional FFTs to distributed data chunks, then performs a redistribution to align the data for the next transform direction.

The algorithm operates on three distributed arrays $A$, $B$, and $C$, representing the layouts required for the x-, y-, and z-direction transforms. Each stage applies local one-dimensional FFTs along the contiguous dimension of its layout, with data redistributed between arrays to align with the next transform direction. This design avoids a single static decomposition and improves data locality across stages.

\begin{figure}[t]
\vspace{-0.3cm}
\begin{algorithm}[H]
\small
\caption{3D FFT with Pencil Decomposition}
\label{alg:fft_pencil}
\begin{algorithmic}[1]
\vspace{-0.2cm}
\Require 3D dataset $A$ of size $(N_x, N_y, N_z)$ distributed into pencils
\Ensure Transformed 3D dataset $A$ in Fourier space

\State Initialize three distribution patterns:
\State $D_1 \gets$ distribution with complete $x$-dimension (allocated)
\State $D_2 \gets$ distribution with complete $y$-dimension (zeros)
\State $D_3 \gets$ distribution with complete $z$-dimension (zeros)

\State \textbf{Stage 1: FFT along $x$-dimension (local)}
\ForAll{local chunks of $A$ (with distribution $D_1$)}
    \State \texttt{Spawn} a task to apply 1D FFT along $x$-dimension
\EndFor

\State \Call{redistribute\_chunks!}{$B$, $A$} 
\State Allocate distributed array $B$ with distribution $D_2$

\State \textbf{Stage 2: FFT along $y$-dimension}
\ForAll{local chunks of $B$}
    \State \texttt{Spawn} a task to apply 1D FFT along $y$-dimension
\EndFor

\State \Call{redistribute\_chunks!}{$C$, $B$}
\State Allocate distributed array $C$ with distribution $D_3$

\State \textbf{Stage 3: FFT along $z$-dimension}
\ForAll{local chunks of $C$}
    \State \texttt{Spawn} a task to apply 1D FFT along $z$-dimension
\EndFor

\State Collect the final result from $C$ \texttt{DArray}
\end{algorithmic}
\end{algorithm}
\vspace{-0.9cm}
\end{figure}

Each FFT stage is launched within a \texttt{SpawnDataDeps} block, allowing the runtime scheduler to assign FFT tasks to the processors that already hold the corresponding data chunks.
The \texttt{applyfft} function internally manages cached \texttt{CUDA.cuFFT} or \texttt{FFTW} plans to avoid repeated planning overhead and to reuse FFT plans across similar transforms.

Redistribution calls (lines 9 and 15 in Algorithm~\ref{alg:fft_pencil}) perform distributed transposes to reorganize the data layout for the next FFT direction. For inverse 3D FFTs, the same sequence is applied in reverse order to reconstruct the data from frequency space back to the spatial domain. The inverse process first performs one-dimensional inverse transforms along the final FFT direction, followed by successive redistributions and inverse transforms along the remaining dimensions. By mirroring the forward algorithm, the inverse FFT benefits from the same task scheduling, overlap, and locality optimizations.

\subsection{Asynchronous Redistribution}

The redistribution routine in \textit{DaggerGPUFFTs} implements a fully asynchronous communication procedure based on non-blocking MPI and CUDA streams. The goal is to overlap five stages of work: (1) chunk caching, (2) posting of receives, (3) packing and sending, (4) local copies, and (5) progressive unpacking. 
Algorithm~\ref{alg:flat_redistribute} outlines this process.

\begin{figure}[t]
\begin{algorithm}[H]
\vspace{-0.2cm}
\small
\caption{Asynchronous Redistribution}
\label{alg:flat_redistribute}
\begin{algorithmic}[1]
\Require Distributed arrays src, dst, workspace $W$
\Ensure Data from src redistributed into dst

\State \textbf{Phase 1: Cache local chunks}
\ForAll{chunks owned by local rank}
  \State Fetch chunk into $W.\text{cache}$
\EndFor

\State \textbf{Phase 2: Post receives early}
\ForAll{$r \in W.\text{recv\_ranks}$}
  \State $W.\text{recv\_reqs}[r] \gets \text{MPI.Irecv!}$
\EndFor

\State \textbf{Phase 3: Pack and send}
\ForAll{$r \in W.\text{send\_ranks}$}
  \State Pack data for peer $r$ on CUDA stream
  \State $W.\text{send\_reqs}[r] \gets \text{MPI.Isend}$
\EndFor

\State \textbf{Phase 4: Local copies}
\ForAll{$p \in W.\text{local\_patterns}$}
  \State Launch local copy kernel
\EndFor

\State \textbf{Phase 5: Progressive unpack of receives}
\State $R \gets W.\text{recv\_ranks}$
\While{$R \neq \emptyset$}
  \ForAll{$r \in R$}
    \If{$\text{MPI.Test!}(W.\text{recv\_reqs}[r])$ succeeds}
      \State Unpack received buffer on CUDA stream
      \State $R \gets R \setminus \{r\}$
    \EndIf
  \EndFor
  \If{$R \neq \emptyset$} yield() \EndIf
\EndWhile
\end{algorithmic}
\end{algorithm}
\vspace{-0.9cm}
\end{figure}

\textbf{Phase~1} fetches all chunks assigned to the local rank into GPU memory to minimize future latency.

\textbf{Phase~2} posts non-blocking receives (\texttt{MPI.Irecv!}) for all expected messages. This ensures that receive buffers are available before senders initiate communication,
allowing MPI to process data transfers in the background.

\textbf{Phase~3} packs outgoing data for each peer into contiguous GPU buffers on dedicated CUDA streams and immediately initiates non-blocking sends (\texttt{MPI.Isend}). CUDA-aware MPI handles synchronization automatically and enables concurrent packing and transfers.

\textbf{Phase~4} performs local memory copies between overlapping regions on the same rank.

\textbf{Phase~5} continuously tests for completed receives using \texttt{MPI.Test!} and launches GPU unpacking kernels as soon as messages arrive.

Overall, this redistribution algorithm eliminates global synchronization and achieves fine-grained concurrency across ranks and devices. The CPU implementation follows the same asynchronous redistribution pattern, overlapping communication and local operations to maintain continuous progress, executed entirely in host memory.

\subsection{Asynchronous Task Scheduling}
As discussed in Section~\ref{sec:runtime_model} Equation~(\ref{eq:scheduling_cost}), the runtime scheduler assigns FFT tasks based on a combined estimate of computation and communication costs. Each FFT stage generates a large number of fine-grained tasks that operate on distributed data chunks, which must be placed across heterogeneous workers to achieve both locality and balance.

The scheduling procedure consists of two phases, as shown in Algorithm~\ref{alg:locality_scheduler}. In the placement phase, each task is assigned to the worker that maximizes data affinity (i.e., the worker that already holds the majority of its input data). Within that worker, if multiple execution units (e.g., CPU threads or GPU devices) are available, the task is placed on the least-loaded unit according to the current runtime load state. In the correction phase, the scheduler evaluates the load distribution across all workers. If the variance in estimated loads exceeds a configurable threshold, a rebalancing step migrates pending tasks from overloaded workers to underutilized ones.

This two-level strategy enables fully asynchronous execution within each FFT stage. Adaptive load adjustments ensure better balance after redistributions and on systems with mixed hardware resources. The scheduler operates continuously without global barriers, allowing FFT tasks to progress independently as soon as their dependencies are satisfied.

\begin{figure}[t]
\begin{algorithm}[H]
\vspace{-0.2cm}
\small
\caption{Locality-Aware Dynamic Task Scheduler}
\label{alg:locality_scheduler}
\begin{algorithmic}[1]
\Require Task set $T$, worker set $W$, load states $L$
\Ensure Assignment $\sigma : T \rightarrow W$
\For{each task $t \in T$}
    \State $w^\ast \gets \arg\max_w \text{Affinity}(t,w)$
    \State $\sigma(t) \gets w^\ast$
    \State $L[w^\ast] \gets L[w^\ast] + \Call{EstimateCost}{t,w^\ast}$
\EndFor
\If{$\Call{Variance}(L) > \text{THRESHOLD}$}
    \State $\sigma \gets \Call{Rebalance}{\sigma,W,L}$
\EndIf
\State \Return $\sigma$
\end{algorithmic}
\end{algorithm}
\vspace{-0.9cm}
\end{figure}

As summarized in Algorithm~\ref{alg:locality_scheduler}, this approach runs in $O(|T| \cdot W)$ time, where $W$ is the number of candidate workers evaluated per task. In practice, $W = O(1)$ since each task requires data from only adjacent workers in the domain decomposition, yielding effective linear complexity $O(|T|)$. By combining data-affinity-driven placement with dynamic load correction, the scheduler maintains high concurrency while reducing inter-worker communication, forming the foundation of \textit{DaggerFFT}'s asynchronous execution model.

\vspace{-0.1cm}

\subsection{Data Dependency Scheduler for FFT Tasks}

The original \textit{Dagger.jl}\cite{b24} runtime provides a general-purpose \texttt{DataDeps} mechanism for managing read/write dependencies across arbitrary tasks. While useful in workloads with deeply nested or interdependent computations, we realized that its fine-grained aliasing analysis introduces nontrivial overhead for structured and highly parallel computations such as local FFTs, where data access patterns are regular and non-aliasing by design. To address this, we implemented a data dependency tracker that operates on top of Dagger's runtime system, optimized for these kinds of workloads. This \texttt{DataDepsTaskQueue} maintains per-chunk read/write tracking rather than global aliasing graphs, enabling immediate dispatch of independent chunk-level FFT tasks once their dependencies are satisfied. This approach transforms FFT execution into a streaming pipeline that maintains per-processor load estimates, integrates rank-level information from the MPI decomposition plan, and implements dynamic task placement on the least-loaded workers. This design significantly reduces global synchronization and improves throughput without requiring any changes to \textit{Dagger}’s task abstraction or user-facing API.

\begin{figure}[t]
\vspace{-0.2cm}
\centering
\includegraphics[width=0.8\linewidth]{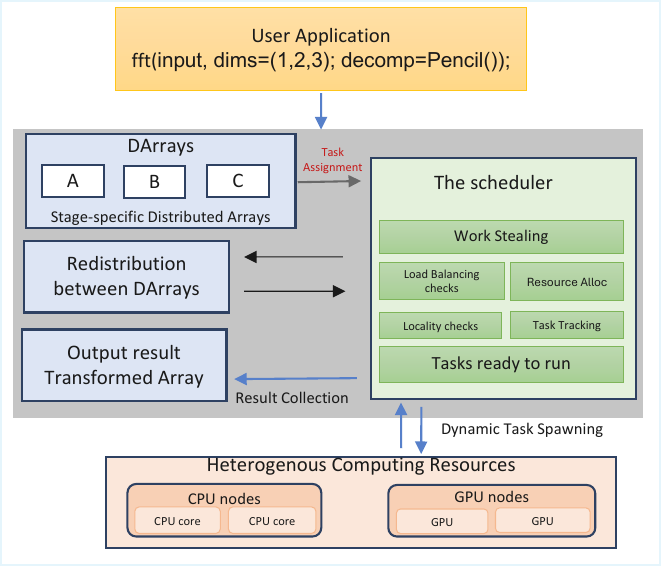}
    %\hline
    %\vspace{-0.5cm}
    \caption{Complete distributed FFT Framework, from a user's view, for resource assignment. The system dynamically constructs a task graph over stage-specific distributed arrays and maps tasks to computing resources using the scheduler.}
    \vspace{-0.4cm}
    \label{fig:framework}
\end{figure}

\section{Implementation}

\subsection{DaggerFFT Framework}
To make our FFT framework accessible and user-friendly, we expose a unified public API through the \textit{DaggerFFT} interface. Users can invoke distributed FFT computations with minimal code changes by calling \texttt{fft} on standard Julia arrays, optionally specifying the desired decomposition strategy and transform type.

\textit{DaggerFFT} supports all standard transform types, including real-to-complex (R2C), complex-to-complex (C2C), and real-to-real (R2R) FFTs, as well as one-, two-, and three-dimensional transforms under both slab and pencil decompositions. Internally, each FFT call triggers a backend-aware execution plan that dynamically allocates \texttt{DArrays}, applies transforms in a task-parallel manner, and handles inter-stage communication using asynchronous redistribution operations. Figure~\ref{fig:framework} illustrates the overall architecture of \textit{DaggerFFT}. Each user-level \texttt{fft} call is decomposed into distributed \texttt{DArrays} that flow through successive FFT stages (A–C). A dynamic scheduler manages task spawning, dependency tracking, and load balancing across hardware resources. Redistribution operations connect these stages asynchronously.
In this paper, we only go through a detailed scenario for complex-to-complex 3D FFTs and cover the algorithmic steps for a pencil decomposition. The other variants follow the same scheduling and communication principles.

\subsection{The Plan Creation and Application}
\label{subsec:plan}

Our \textit{DaggerFFT} framework is designed to perform Fast Fourier Transforms efficiently across both CPU and GPU platforms. The framework employs specialized structs to encapsulate the parameters and metadata required for different types of FFT operations. We support Complex-to-Complex (FFT) transforms, where both input and output are complex-valued arrays; RFFT (Real-to-Complex) transforms optimized for real-valued data, which reduce memory overhead by exploiting Hermitian symmetry\cite{b8}; and Real-to-Real (R2R) transformations such as the discrete cosine transform (DCT) and discrete sine transform (DST).~\cite{b9}

The execution of FFTs follows a plan-based approach inspired by established frameworks such as \textit{FFTW}\cite{b1} and \textit{cuFFT}\cite{b11}. Each FFT computation starts by creating a reusable plan that determines the best execution strategy based on the data size, dimensionality, and transformation type. Once created, plans are applied to data chunks.
To minimize repeated planning overhead, \textit{DaggerFFT} employs a plan caching mechanism that stores and reuses previously generated FFT plans. When an FFT operation is requested, the framework calls a lookup routine that constructs a unique cache key based on the data type, array size, transform type, and transform dimension tuple. If a matching plan is found in the cache, it is retrieved and reused; otherwise, a new plan is created using the appropriate backend (e.g., \texttt{CUDA.CUFFT.plan\_fft}) and inserted into the cache for future use. This process is handled transparently by the function \texttt{get\_or\_create\_plan}, which ensures that plan creation is performed only once per distinct transform configuration.

This caching strategy significantly reduces runtime overhead during large distributed FFTs where many identical transforms are applied across different chunks. Since each chunk typically shares the same size and transform parameters, the same plan can be safely reused across tasks and GPU streams. The result is improved performance consistency, reduced kernel launch latency, and lower memory allocation overhead during multi-stage FFT execution.

\begin{figure}[t]
\centering
\begin{threeparttable}
\captionsetup{type=table}
\caption{Effect of the scheduling runtime on the first FFT stage ($512^3$ grid, 16~ranks, one per core).}
\label{tab:scheduler_local}

\normalsize
\begin{tabular}{l>{\centering\arraybackslash}p{2.8cm}>{\centering\arraybackslash}p{2.8cm}}
\toprule
Decomposition & SimpleMPIFFT (s) & DaggerFFT (s) \\
\midrule
Pencil (1D FFT) & 0.040 & 0.026 \\
Slab (2D FFT)   & 0.100 & 0.060 \\
\bottomrule
\end{tabular}
\end{threeparttable}
\vspace{-0.4cm}
\end{figure}

\subsection{Impact of the Scheduling Runtime on Local FFT Stages}
Before analyzing global pipeline behavior, we first isolate the contribution of \textit{DaggerFFT}'s task scheduler on local FFT execution. To do this, we compare the runtime of a single FFT stage executed under the Dagger scheduling runtime against an equivalent static distributed FFT implementation written in plain Julia with explicit MPI communication and identical local FFT kernels (SimpleMPIFFT). Both implementations use cached \textit{FFTW}\cite{b1} plans to ensure fair comparison of scheduling benefits.

Table~\ref{tab:scheduler_local} reports execution times for the first FFT stage of both pencil (stage 1 of Algorithm\ref{alg:fft_pencil}) and slab decompositions on a $512^3$ grid using 16~MPI~ranks. The \textit{DaggerFFT} implementation achieves substantially lower runtime, up to $1.5\times$ faster for pencil and $1.7\times$ for slab decompositions. The performance benefit is more pronounced in the slab decomposition, where each task performs a more compute-intensive 2D FFT and allows the scheduler to better exploit task parallelism and overlap.

This performance gap stems from \textit{DaggerFFT}'s asynchronous scheduling and dependency management. While both implementations apply FFTs independently per chunk, SimpleMPIFFT enforces implicit synchronization via blocking loops. In contrast, \textit{DaggerFFT} expresses each chunk as an independent task in a dependency graph, enabling immediate dispatch as soon as inputs are ready. This eliminates unnecessary barriers and idle time, yielding measurable gains even in local-only FFT stages without communication.

\subsection{Work stealing behavior}
\begin{figure}[t]
\begin{threeparttable}
    \centering
    \captionsetup{type=table}
    \caption{Effect of work stealing under load imbalance.}
    \label{tab:workstealing_imbalance}

    \begin{tabular}{lcc}
    \toprule
    Metric & Stealing OFF & Stealing ON \\
    \midrule
    Total time (s)              & $\approx 8.9$ & $\approx 8.6$ \\
    Imbalance (\%$)^{*}$              & $\approx 45$ & $\approx 10$ \\
    Max / Min thread time (s)   & $\approx 8.9 / 2.0$ & $\approx 8.6 / 7.8$ \\
    Avg per-thread tasks        & 4.0 & 4.0 \\
    \bottomrule
    \end{tabular}
    \begin{tablenotes}
    \footnotesize
    \item[*] Imbalance computed as $\mathrm{std}(\text{per-thread times}) / \mathrm{mean} \times 100\%$.
    \end{tablenotes}
    \end{threeparttable}

    \vspace{0.4cm}

    \captionsetup{type=figure}
    \includegraphics[width=1.0\linewidth]{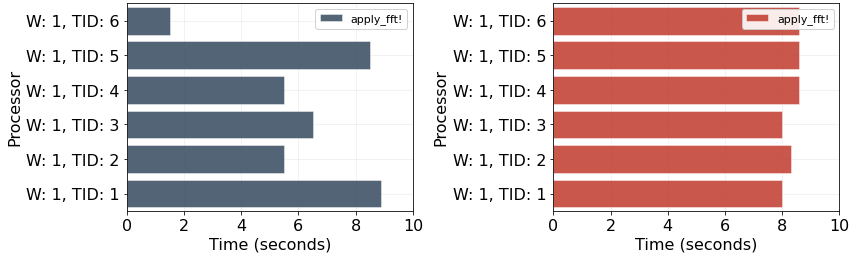}
    %\hfill

    \caption{Execution timelines for applying a 1D FFT (Stage~1 in Algorithm~\ref{alg:fft_pencil}) across six parallel threads (workers) on a single node, with (right) and without (left) work stealing enabled.}

    \label{fig:scheduling_comparison}
    \vspace{-0.5cm}
\end{figure}

To isolate the impact of dynamic load balancing, we introduced an artificial imbalance by assigning a subset of larger data chunks to specific threads. Table~\ref{tab:workstealing_imbalance} summarizes the resulting per-thread statistics with work stealing disabled and enabled.
Without stealing, several threads completed early while others processed heavier workloads, yielding a $\sim$45\% runtime imbalance and an overall execution time of approximately 8.9~s. When work stealing was enabled, idle threads opportunistically executed remaining tasks from overloaded peers, reducing the imbalance to about 10\% and improving the total runtime to 8.6~s. Importantly, this measured execution time already includes the scheduler’s overhead for task monitoring and redistribution and indicates that the benefits of dynamic balancing outweigh its runtime cost in this case.

This controlled experiment demonstrates that the scheduler can effectively mitigate temporary imbalance through opportunistic task redistribution.
However, in typical distributed FFT workloads, the dominant performance improvements arise from other runtime mechanisms, such as asynchronous task spawning, efficient resource utilization, and memory reuse, while work stealing primarily acts as a corrective mechanism under highly irregular load conditions. Together, these features enable smooth utilization of available cores and devices, particularly when chunk processing costs vary across ranks or hardware backends.

\section{Experimental Evaluation}
\subsection{Experimental Setup}
We evaluated our framework on clusters equipped with CPUs and GPUs. Tables~\ref{tab:hardware1} and~\ref{tab:software} summarize the hardware and software platforms used.

\begin{figure}[t]
\centering

\begin{threeparttable}
\captionsetup{type=table}
\caption{CPU and NVIDIA GPU details.}
\label{tab:hardware1}

\normalsize
\begin{tabular}{lp{2.8cm}p{2.8cm}}
\toprule
Component & CPU & NVIDIA GPU \\
\midrule
Processor & Intel Xeon Gold 6240R & Tesla V100-SXM2 \\
Cores/Units & 48 cores & 4 GPUs/node \\
Clock Speed & 2.4\,GHz base & -- \\
Memory & 36\,MB L3 cache & 32\,GB HBM2 \\
Bandwidth & -- & 900\,GB/s \\
Interconnect & InfiniBand HDR & NVLink \\
\bottomrule
\end{tabular}
\end{threeparttable}

\vspace{0.2cm} 
\begin{threeparttable}
\captionsetup{type=table}
\caption{Software configuration.}
\label{tab:software}

\normalsize
\begin{tabular}{lp{5.5cm}}
\toprule
Component & Version/Details \\
\midrule
Julia & v1.11.7 \\
FFT Libraries & FFTW~\cite{b21} v3.3.10, cuFFT \\
GPU Software & CUDA 12.3, CUDA.jl v5.5.2~\cite{b18} \\
MPI & Open MPI v4.1.6, MPI.jl~\cite{b22} v0.20.22 \\
Task Scheduling & Dagger.jl~\cite{b23} v0.18.14 \\
\bottomrule
\end{tabular}
\end{threeparttable}

\vspace{-0.5cm}
\end{figure}

\label{subsec:cpu_comparison}
\begin{figure*}[t]
    %\vspace{-0.5cm}
    \centering
    \includegraphics[width=\linewidth]{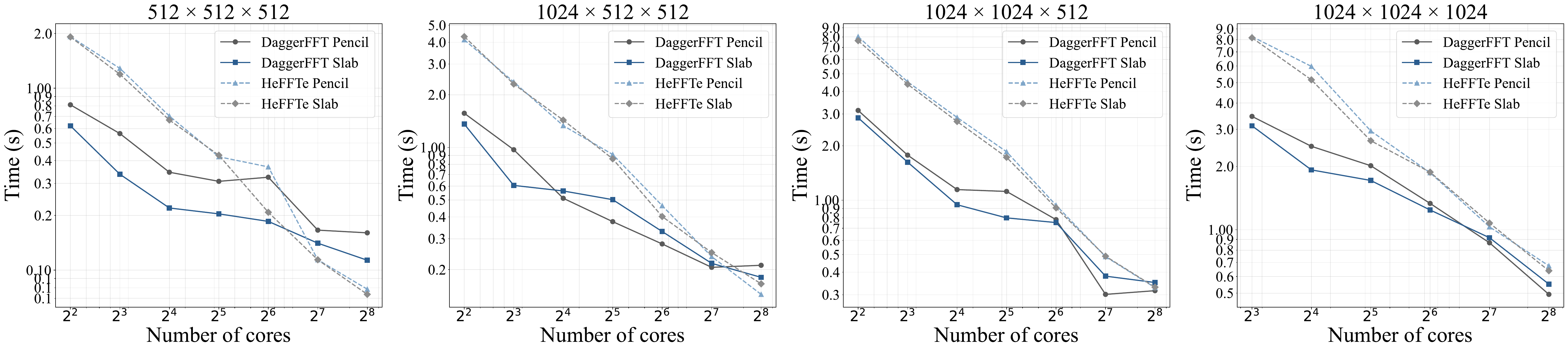}
    \caption{Strong-scaling comparison of three-dimensional single-precision (FP32) distributed FFTs between \textit{DaggerFFT} and \textit{heFFTe}, using both pencil (2D) and slab (1D) decompositions across multiple grid sizes. Each rank is mapped to a single CPU core. Both libraries employ \textit{FFTW} as the local transform backend.}
    \label{fig:cpucomparison}
    \vspace{-0.3cm}
\end{figure*}

\begin{figure*}[h]
    \centering
    \includegraphics[width=\linewidth]{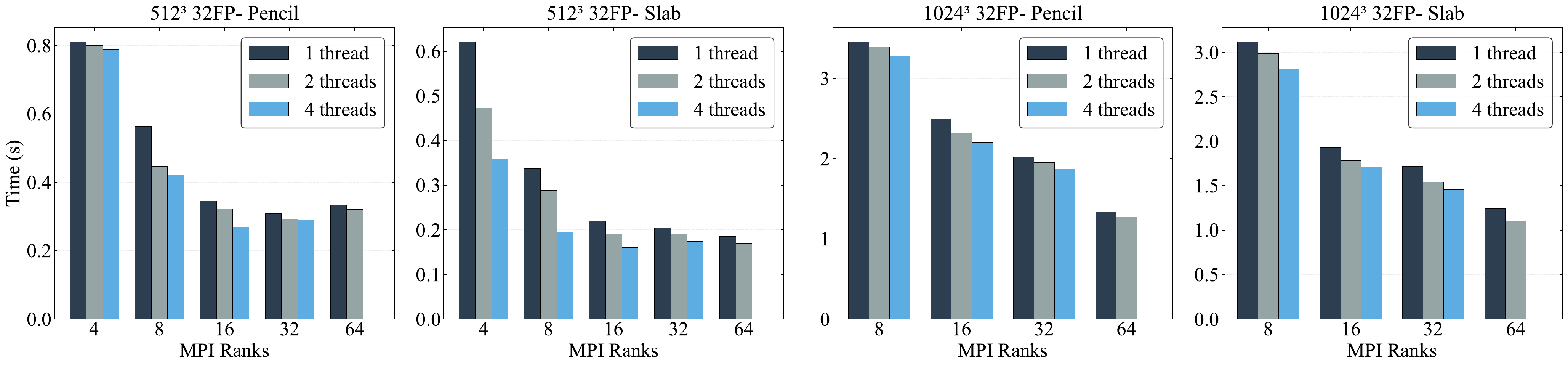}
    %\hline
    \vspace{-0.5cm}
    \caption{Performance of hybrid MPI with threading for three-dimensional FP32 FFTs using slab and pencil decompositions. Each bar shows total runtime for different thread counts per rank, with each thread assigned to a dedicated CPU core.}
    \vspace{-0.3cm}

    \label{fig:threading}
\end{figure*}

The performance of the distributed FFT was assessed through a series of scalability tests and comparisons against existing FFT libraries on multi-core CPU and multi-GPU systems. Following common benchmarking practices for evaluating distributed FFT libraries, we report on the time spent on the three FFT transforms and the two data redistribution steps, corresponding to stages 1 through 3 in Algorithm~\ref{alg:fft_pencil}. We report the average execution time over multiple iterations to ensure consistency and reduce the impact of runtime fluctuations. Even though different transform types are supported in the framework, all the results and comparisons are on Complex-to-Complex (C2C) FFT transforms and Point-To-Point MPI for communication, which is implemented according to Algorithm~\ref{alg:flat_redistribute}. For CPU-based transforms, we use the \textit{FFTW} library as backend via \emph{FFTW.jl}~\cite{b21}, while GPU-based transforms are executed using \textit{cuFFT} through \emph{CUDA.jl}~\cite{b18}.

\subsubsection{Scalability Analysis}
\label{subsec:scaling}

Figure~\ref{fig:cpucomparison} presents the strong scaling behavior of \textit{DaggerFFT} for three-dimensional complex-to-complex single-precision (32-bit floating-point) transforms using both 2D and 1D decompositions. Each data point corresponds to the total wall-clock time for a complete 3D forward and backward FFT, including local transforms and redistribution phases, across an increasing number of CPU cores, with one rank running on a dedicated core.

\textit{DaggerFFT} demonstrates sub-linear strong scaling across all tested configurations, with parallel efficiency improving substantially as problem size increases. For the smallest grid ($512^3$), slab decomposition achieves a $5.5\times$ speedup from 4 to 256 ranks, marginally outperforming pencil's $5.1\times$ speedup.
As problem size increases, scaling efficiency improves. The largest problem ($1024^3$) exhibits the best scaling characteristics. Pencil decomposition achieves $7.0\times$ speedup from 8 to 256 ranks, outperforming slab's $5.7\times$ speedup. Overhead limits scalability for smaller problem sizes (e.g., $512^3$ plateaus beyond 128 ranks), but larger grids ($1024^3$) maintain stronger scaling due to improved computation-to-communication ratios that better amortize redistribution costs. These results confirm that \textit{DaggerFFT}'s asynchronous redistribution and task-based scheduling effectively utilize resources while avoiding bulk-synchronous penalties. Strong scaling efficiency follows the expected trend, where larger workloads per rank better amortize communication and scheduler overhead.

\begin{figure*}[t]
    \centering
   % \vspace{-0.2cm}
    \includegraphics[width=1.0\linewidth]{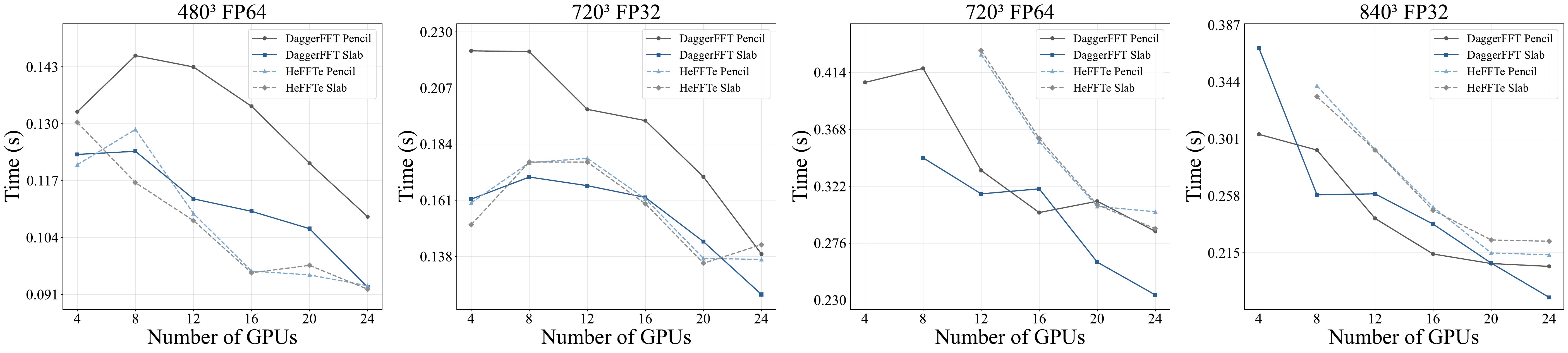}
    \caption{Strong-scaling comparison of three-dimensional FP32 and FP64 FFTs between \textit{DaggerFFT} and \textit{heFFTe}, using both pencil (2D) and slab (1D) decompositions across multiple grid sizes. Each rank is mapped to a GPU. Both libraries employ \textit{cuFFT} as the local transform backend.}
    \label{fig:gpu_compare}
    \vspace{-0.5cm}
\end{figure*}

\subsubsection{Hybrid MPI + Threading Analysis}
A key advantage of \textit{DaggerFFT} is its ability to transparently support concurrent execution across both distributed MPI ranks and shared-memory threads within each rank, with each thread pinned to a dedicated physical core. The task-based runtime automatically decomposes local FFT operations into parallel work units distributed across available threads, providing an additional level of parallelism on multi-core nodes without modifying the user-facing API.

Figure~\ref{fig:threading} demonstrates the impact of intra-rank threading on execution time across both decomposition strategies. Slab decomposition exhibits substantially stronger threading efficiency, particularly for the smaller $512^3$ configuration, where 4-thread execution achieves $1.50\times$ average speedup compared to pencil's $1.18\times$ speedup. This 27\% threading advantage arises because each slab stage performs a more computationally intensive 2D FFT per task and threads exploit greater parallel work within each local transform.

In contrast, pencil decomposition divides the domain more finely across available ranks, resulting in smaller per-rank workloads with limited parallelism available for threading. For the larger $1024^3$ problem, this difference diminishes, as both decompositions provide enough local work to fully utilize the available cores. These results indicate that hybrid parallelism is more effective for slab-based configurations when the per-rank problem size is modest.

\subsubsection{GPU Scalability Analysis}
Figure~\ref{fig:gpu_compare} presents the strong-scaling performance of \textit{DaggerFFT} on multi-GPU systems using the \textit{cuFFT} backend across varying problem sizes and floating-point precisions. Each plot shows execution time for a complete 3D FFT as GPU count increases from 4 to 24.

\textit{DaggerFFT} exhibits consistent strong-scaling behavior across all configurations, with scaling efficiency improving substantially for larger problem sizes. The largest configuration tested ($840^3$ FP32 with slab decomposition) achieves the best overall performance, reducing execution time from 0.370\,s on 4 GPUs to 0.181\,s on 24 GPUs (2.04$\times$ speedup. For double-precision workloads, the $720^3$ FP64 pencil configuration demonstrates a 1.42$\times$ speedup across the same GPU range.

Smaller problem sizes exhibit more limited scaling gains due to communication overhead. For instance, the $480^3$ FP64 pencil configuration achieves only 1.22$\times$ speedup, as inter-GPU data transfers and kernel launch latencies consume a larger fraction of total runtime when per-GPU workloads are modest. However, as grid size increases, the computation-to-communication ratio improves, enabling \textit{DaggerFFT} to better exploit parallelism through overlapped asynchronous transfers and fine-grained task scheduling that maintains high device utilization throughout the FFT implementation. The results confirm that multi-GPU strong scaling becomes increasingly effective as problem size grows, where larger per-GPU workloads better amortize communication and distribution costs.

\subsubsection{CPU Performance Evaluation against Baseline Libraries}
To evaluate \textit{DaggerFFT}'s performance on CPU clusters, we compare against \textit{heFFTe}~\cite{b5}, a modern C++ framework that represents the current state of the art in scalable distributed FFTs. Both libraries use \textit{FFTW}~\cite{b1} as the local backend. Figure~\ref{fig:cpucomparison} presents strong-scaling results for several three-dimensional problem sizes in single precision.

\textit{DaggerFFT} consistently outperforms \textit{heFFTe} across all configurations, with the performance advantage most pronounced at low to moderate process counts where computation still dominates communication overhead. At 4 ranks, \textit{DaggerFFT} achieves 2.37$\times$ speedup for $512^3$ pencil and 2.67$\times$ for $1024\times512^2$ slab decomposition. As problem size increases, peak speedups occur at higher rank counts: for $1024^3$, \textit{DaggerFFT} achieves 2.40$\times$ (pencil) and 2.68$\times$ (slab) at 16 ranks.

The performance advantage diminishes at very high process counts where scheduler costs dominate. For larger grids ($1024^3$), \textit{DaggerFFT} maintains a 1.16$-$1.37$\times$ advantage even at 256 ranks, demonstrating that its task-based approach scales more effectively when per-rank workloads remain substantial.

\subsubsection{GPU Performance Evaluation against Baseline Libraries}
\label{subsec:gpu_comparison}
To evaluate \textit{DaggerFFT}'s performance on GPU clusters, we compare against \textit{heFFTe}, using \textit{cuFFT} as the local backend for both frameworks. The performance comparison shows  as problem size increases, \textit{DaggerFFT}'s performance advantage emerges. For $840^3$ FP32, \textit{DaggerFFT} outperforms \textit{heFFTe} by 1.04$-$1.35$\times$ across all configurations, with slab decomposition showing the strongest gains. The $720^3$ FP64 configurations demonstrate intermediate behavior, as \textit{DaggerFFT} achieves 1.13$-$1.36$\times$ speedup for most GPU counts.

The results indicate that \textit{DaggerFFT}'s task-based asynchronous execution model effectively hides latency for larger workloads in GPU backend, where the per-GPU computational intensity is sufficient to amortize scheduling overhead. At smaller scales, the fixed costs of dynamic task management exceed the benefits of overlap, which makes static execution model more efficient. 

\subsection{Application to Oceananigans Poisson Solver}
\label{subsec:oceananigans_poisson}

To demonstrate the true performance impact of our distributed FFT framework on a full-scale HPC application, we integrated \textit{DaggerGPUFFTs} into \textit{Oceananigans.jl}\cite{b27}, a high-performance Julia package for simulating fluid dynamics in ocean and climate modeling. \textit{Oceananigans} is widely used by the geophysical fluid dynamics community because it offers a flexible, GPU-accelerated environment for large-scale simulations of turbulence and stratified flows. One of the central challenges in such simulations is the efficient solution of the pressure Poisson equation~\cite{b33}, which enforces incompressibility at each timestep. This equation typically accounts for a significant fraction of the total runtime, making it a target for performance optimization.

In \textit{Oceananigans}, the Poisson equation is solved using FFT-based methods whenever periodic-boundary conditions are present. For large-scale ocean models requiring billions of grid points, a distributed FFT library that provides the combination of scalability, performance, and ease of integration is required.
\begin{figure}[t]
%\vspace{-0.2cm}
    \centering
    \includegraphics[width=1.0\linewidth]{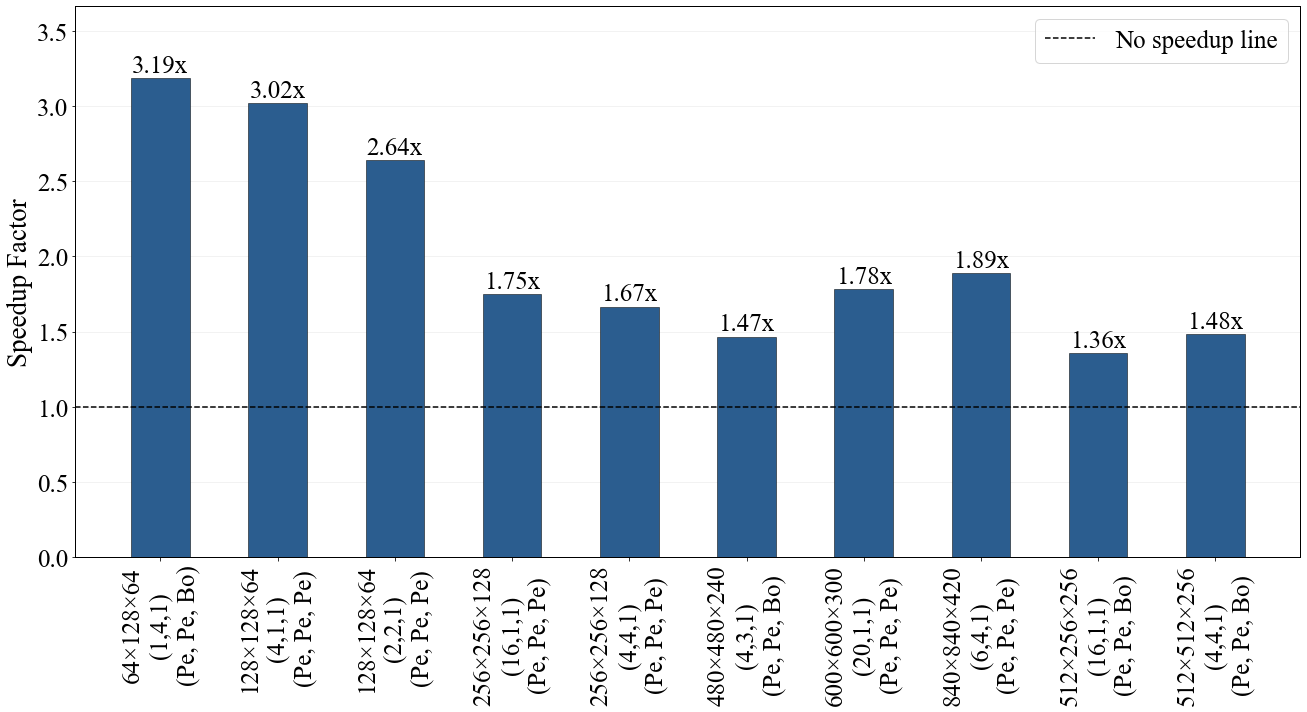}
    \caption{Speedup of the Oceananigans Poisson solver with \textit{DaggerGPUFFTs} compared to the baseline solver. Each rank is assigned to one GPU.}
   \vspace{-0.5cm}
    \label{fig:oceananigans}
\end{figure}

We replaced \textit{Oceananigans}' native FFT routines in the Poisson solver with our distributed FFT implementation from \textit{DaggerGPUFFTs}. This design enables the solver to achieve scalable performance on heterogeneous systems, while retaining the simplicity and usability of the original API. Figure~\ref{fig:oceananigans} shows the end-to-end performance improvements of the Poisson solver after integrating \textit{DaggerGPUFFTs} with the \texttt{cuFFT} backend for two topologies, which are (Periodic, Periodic, Periodic) and (Periodic, Periodic, Bounded). The distributed FFT pipeline delivers substantial speedups over the original implementation, with peak acceleration of $3.19\times$ observed for smaller problem sizes where computation and memory limitations dominate. Even at larger scales and multiple nodes, the solver consistently achieves $1.3\times$–$1.8\times$ improvements due to the asynchronous redistribution of \texttt{DArrays} and better GPU utilization. These results highlight that the distributed, task-based FFT implementation is not only effective in controlled benchmarks, but also delivers tangible improvements in a widely adopted scientific modeling framework. For scientists, the transition requires no modifications to existing workflows, as the solver preserves the same high-level interface and fits naturally into existing \textit{Oceananigans} simulations. As a result, simulations can now be executed at higher resolutions and larger scales, that enables new opportunities for ocean and climate research.

\section{Discussion}
\subsection{Limitations and Overhead Analysis}
\label{sec:limitations}
As demonstrated in Figure~\ref{fig:cpucomparison}, in smaller problem sizes, the scalability of \textit{DaggerFFT} diminishes as the number of cores (and MPI ranks) highly increases. This behavior stems from the overhead introduced by the dynamic task scheduler, which must track an expanding number of task dependencies, chunk mappings, and scheduling decisions as the domain is decomposed into finer chunks. The costs eventually outweigh the benefits of asynchronous execution for lightweight FFT computations.

\begin{figure}[t]
%\vspace{-0.2cm}
    \centering
    \includegraphics[width=0.85\linewidth]{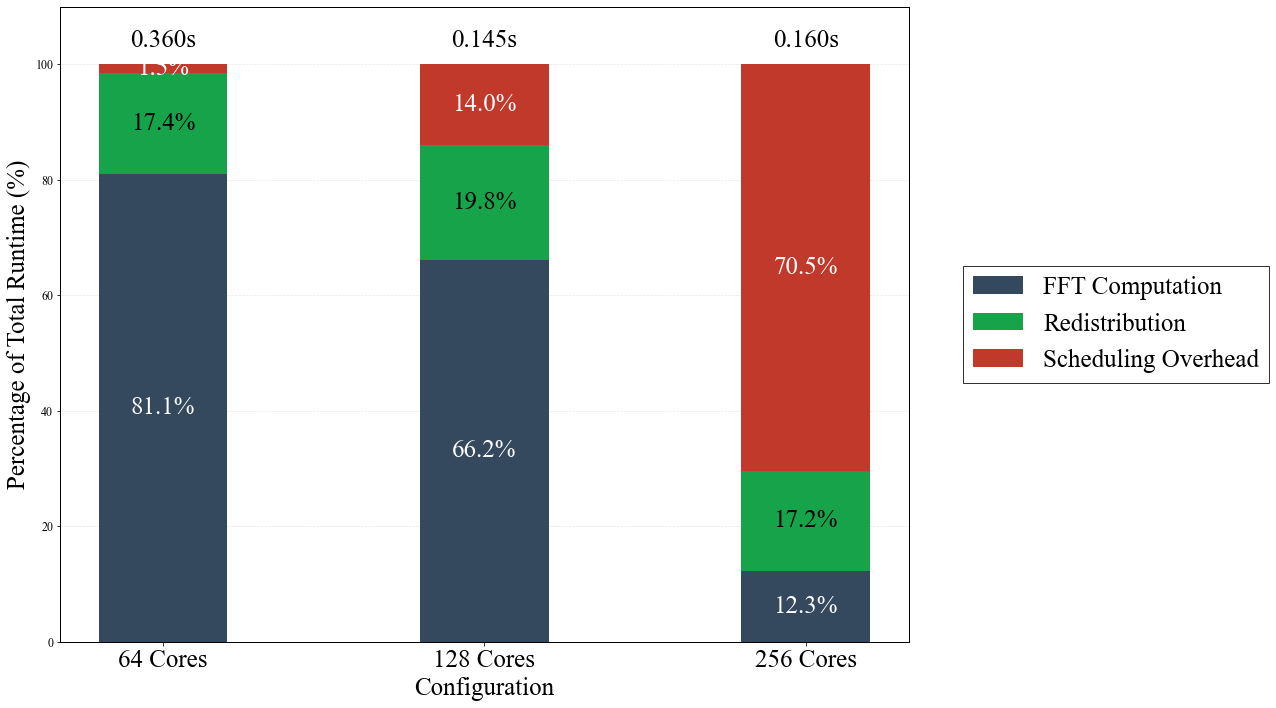}
    \caption{Performance breakdown for 512³ pencil decomposition showing the percentage of total runtime spent in FFT computation, redistribution, and scheduling overhead. Each rank is assigned to one core.}
   \vspace{-0.5cm}
    \label{fig:limitation}
\end{figure}

To quantify this overhead, we profiled the worst-case scenario in Figure~\ref{fig:cpucomparison}: $512^3$ pencil decomposition at high rank counts. Figure~\ref{fig:limitation} presents the performance breakdown, revealing a dramatic shift in time distribution. At 64 ranks (0.36s total), FFT computations are dominant, consuming 81.4\% of the runtime, while redistribution accounts for 17.4\% and scheduling overhead remains negligible at 1.5\%. At 128 ranks (0.145s), the percentage of FFT computations drops to 66.2\%, redistribution increases to 19.8\%, and overhead grows to 14.0\%. However, at 256 ranks (0.16s), the percentage of FFT computation drops to just 12.3\%, redistribution consumes 17.2\%, and scheduling overhead explodes to 70.5\% of total runtime. This breakdown confirms that excessive task decomposition saturates the runtime system with dependency resolution and fine-grained synchronization costs. These results highlight the fundamental trade-off in task-parallel FFT implementations. Asynchronous execution provides substantial benefits when per-task work is sufficient to hide coordination overhead, but cannot be amortized when the size of problem per rank falls below a critical threshold.

\section{Conclusion}
In this paper, we introduced \textit{DaggerFFT}, a distributed FFT framework that expresses FFT computations as dynamically scheduled task graphs. Implemented entirely in the Julia programming language, \textit{DaggerFFT} demonstrates that modern high-level languages can achieve performance competitive with low-level implementations while providing greater flexibility. \textit{DaggerFFT} enables modular and portable FFT execution across CPUs and GPUs by integrating different domain decomposition strategies with asynchronous MPI communication and dynamic task scheduling. The framework employs hybrid parallelism to utilize available compute resources efficiently. \textit{DaggerFFT} achieves up to a 1.35$\times$ speedup over baseline GPU implementations and a 2.7$\times$ speedup over CPU libraries. Unlike traditional FFT frameworks that rely on static execution models, \textit{DaggerFFT} sustains high throughput by dynamically balancing computation at runtime without requiring hardware-specific tuning or custom accelerators. These results highlight the effectiveness of dynamic scheduling as a pathway to improved FFT performance and resource efficiency in modern high-performance computing environments.

\balance

\end{document}